\documentclass[aps,twocolumn,amsmath,amssymb,floatfix,fleqn]{revtex4}
\usepackage{lineno,hyperref}
\usepackage{graphicx}
\usepackage{bm}
\usepackage[german,american]{babel}
\usepackage{bigints}
\usepackage{mathtools, nccmath}
\newcommand{\fig}[1]{Fig.\ \ref{#1}}
\newcommand{\eqn}[1]{Eq.\ \eqref{#1}}

\usepackage{accents}
\newcommand*{\dt}[1]{%
  \accentset{\mbox{\large\bfseries .}}{#1}}

\makeatletter
\DeclareRobustCommand{\iscircle}{\mathord{\mathpalette\is@circle\relax}}
\newcommand\is@circle[2]{%
  \begingroup
  \sbox\z@{\raisebox{\depth}{$\m@th#1\bigcirc$}}%
  \sbox\tw@{$#1\square$}%
  \resizebox{!}{\ht\tw@}{\usebox{\z@}}%
  \endgroup
}
\makeatother

\begin{document}

\title{Interpreting Force Response Patterns of a Mechanically Driven Crystallographic Phase Transition}
\author{Arijit Maitra}
\email{arijit.maitra@bmu.edu.in}

\author{Bipin Singh}
\affiliation{Department of Applied Sciences, School of Engineering and Technology, BML Munjal University, NH 8, 67 KM Milestone, Gurugram, Haryana 122413, India }

\date{\today}

\begin{abstract} 
Mechanically induced crystallographic phase transformation that 
reflects dynamic stress responses of intrinsically stochastic 
nature is a pertinent yet much less well-understood phenomenon.
We focus on understanding the physical significance of 
stochasticity and how it can enable inference of principles 
underlying a crystallographic phase transformation.
For interpreting the mechanical responses, a statistical approach of 
mapping the transformation dynamics to a probabilistic escape of 
crystallographic states defined on a free energy landscape
is shown to reliably explain the patterns of response.
We demonstrate that stochastic responses associated with a 
structural phase transformation
can offer a reliable quantitative tool for unravelling the
energy profile, intrinsic kinetics, and microscopic details 
of solid-to-solid crystallographic transitions.
\end{abstract}

\maketitle

\section*{Introduction}
Advances in nanomechanical instrumentation and MEMS (micro-electro-mechanical system) based devices 
are offering new approaches in the field of evaluation and characterization of materials \cite{Minor2019,Garcia2020}.
Use of these techniques enables precise measurements of how a material responds
to an externally imposed displacement or force ramp, captured in
the form of deformation response-stimulus patterns, e.g., force--displacement or stress--strain correlations.
Nano and micromechanical testing 
of small---sub-micrometre---sized \emph{single-crystalline} solids
have often revealed that the nature of force response patterns is \emph{stochastic} \cite{Dimiduk2005, Dehm2018}.
%
The observed stochasticity 
has a microscopic basis and arises because of intrinsic fluctuations 
in the generative mechanisms  and evolution of 
a small number of imperfections in the crystal lattice
such as twins, stacking faults and dislocations \cite{hosford}. 
We turn to crystalline metallic systems, where nanoscale stochastic force-response patterns 
offer a powerful yet untapped quantitative perspective
of the microscopic mechanisms underlying deformation, which  are
otherwise inaccessible in classical (bulk) testing methods.
An important task of quantitative analyses in 
dynamic force spectroscopy of materials, discussed here,
is to infer the hidden---microscopic---information of deformation rate-processes 
that underlie system level---macroscopic---behaviour.
\begin{figure}
\includegraphics[width=\linewidth,trim= 0mm 0mm 0mm 0mm,clip]{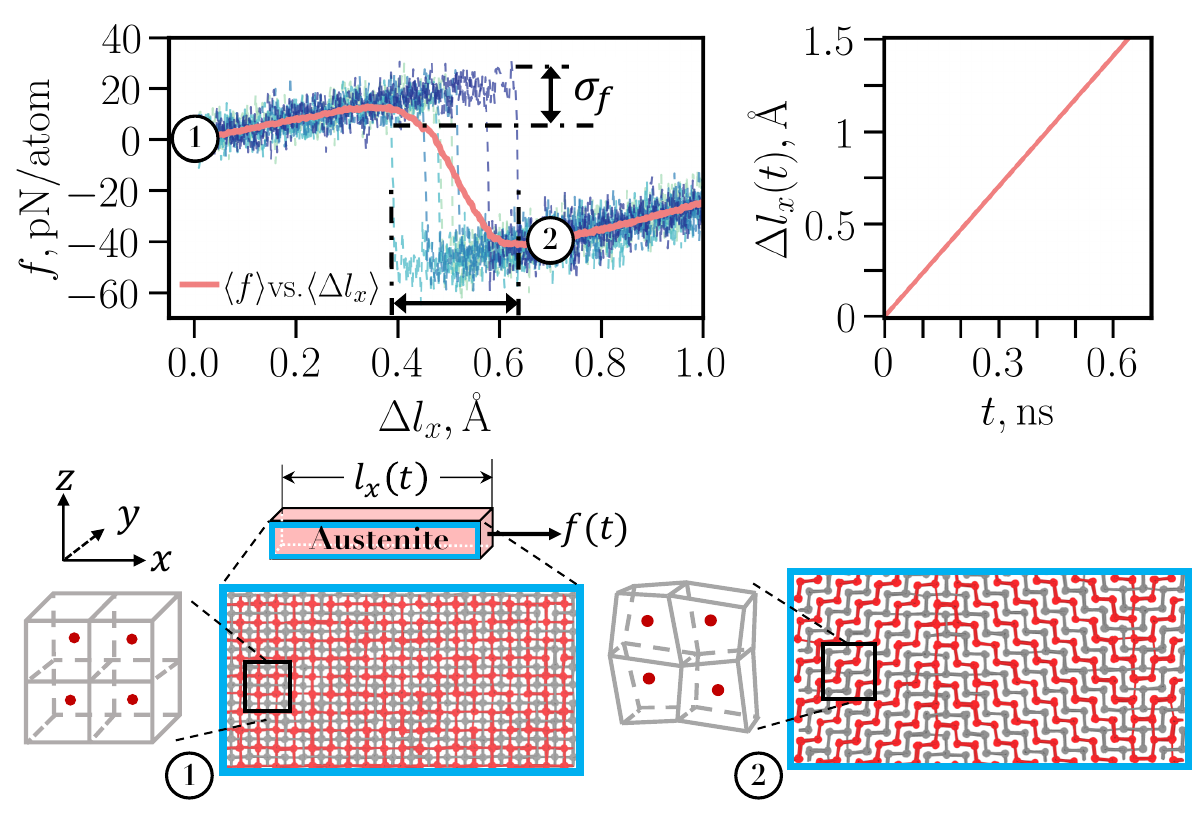}
\caption{ {\bf Stochastic Nature of Phase Transition Response.  \label{fig1}}
\textbf{Top-Left} (- - -) Uniaxial tensile force, $f$, vs.~displacement, $\Delta l_x$, obtained
from independent molecular dynamics simulations, performed under identical conditions: rate of extension  $\dot{l_x}$ = 0.00235 \AA/ps, and temperature T = 300 K. Representative atomic views of titanium nickel alloy structure marked \textcircled{\scriptsize{1}} and \textcircled{\scriptsize{2}} are shown in bottom panels. 
\textbf{Top-Right} Imposed tensile extension, $\Delta l_x$,  vs.~$t$, which produced responses of the Top-Left panel.
\textbf{Bottom-Left} Schematic deformation of austenite along [100], coinciding with the $x$-axis of the simulation box. Snapshot of $xz$ cross-section of the simulation box, unstressed at $t=0$, and zoomed unit cell is shown. Red and grey dots are titanium and nickel atoms. 
\textbf{Bottom-Right} Atomic arrangement and unit cell of martensite at $t$ = 0.32 ns, extracted from a simulation trajectory, after relaxation of the peak force. 
}
\end{figure}

By the term \emph{stochastic}, we refer to a non-deterministic nature 
that is characterizable through a statistical distribution of 
response variables, observed in experiments or simulations, e.g., critical field (force) or time of 
an observed lattice instability.  
To allow the significance of stochasticity interpreted, 
we describe a statistical-mechanical approach for modelling the distribution
that ultimately provides insights into 
a relevant deformation mechanism.
We illustrate the model to learn the characteristics of a \emph{twin} mechanism
operating underneath 
the \emph{pseudoelastic} mechanical behaviour
of a single crystalline titanium nickel alloy \cite{Otsuka2005,Christian1995,Beyerlein2014}.

In our illustration, the stochasticity manifested in the 
ensemble of mechanical responses is due to a 
strain-induced crystallographic \emph{twinning}---a microscopic 
mechanism---implicated in the isothermal solid-to-solid \emph{martensitic} transformation 
of an austenite phase to martensite, 
which in essence involves a change of crystal structure from a body centred cubic to a predominantly monoclinic system.
The stochastic response stems from random and thermally activated microscopic processes, 
as the austenite lattice restructures into a twinned lattice of the martensite.
Apart from fundamental interests, understanding  
stochastic phase transformation-mediated constitutive responses
is important 
for applications and end-use functionalization of materials. For instance, response-fluctuations 
will have an impact on the accuracy and reliability of tiny actuators and micromachines 
that are composed of pseudoelastic and shape memory alloys, a class the titanium nickel alloy 
belongs \cite{Bhattacharya2005,McCracken2020}.

Solving the inference problem, i.e., what mechanistic properties 
underlie the stochasticity of transformation-mediated mechanical responses 
will have a broader impact on the analyses of 
nanomechanical testing data. This, however,
is yet to be explored in metallic materials. 
Theoretical and computational models invoking martensitic transformation and microstructural aspects 
for the treatment of mechanical behaviour can be found in a few insightful studies 
\cite{Olson1975,Falk1980,Achenbach1989,Beyerlein2014,Chen2020}.
Here, we show how the intrinsic kinetic properties, free-energy landscape and the range of microscopic 
interactions can be reliably inferred for a deformation mechanism associated
with an observed  distribution of critical forces.

\begin{figure}
\centerline{\includegraphics[width=0.8\linewidth,trim= 0mm 5mm 0mm 10mm,clip]{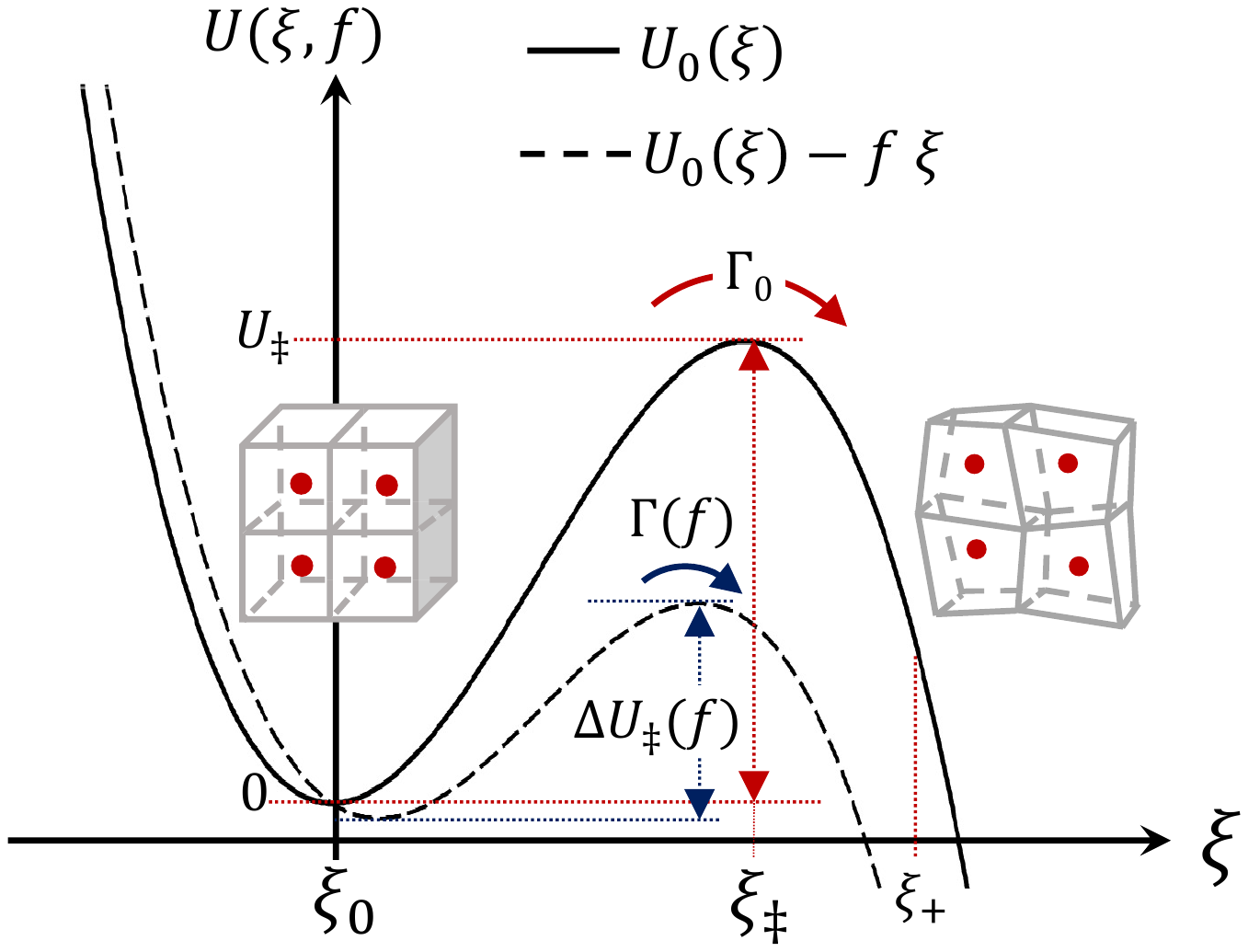} }
\caption{{\bf Energy Profile of a Phase Transition. \label{fig2} } (Solid line) No applied force, $f$=0; and (dashed line) under applied force, $f>0$. Austenite phase is associated with a range of the order parameter $\xi \in (-\infty, \, \xi_\ddag ]$ and martensite to $\xi \in (\xi_\ddag , \, +\infty )  $.}
\end{figure}

Interpretation of stochastic force responses is based on three-fold steps.
First, a distribution of a feature variable, which is to be modelled,
e.g., \emph{time} of a force-response curve to flip abruptly, 
or \emph{critical force} to generate a lattice twin is required; refer \fig{fig1}(Top-Left). 
In our simulated responses, such recognizable features correlate with
the inception of a defect (e.g., twin) mechanism
within an initially defect-free (parent austenite) lattice. 
Second, a model of free-energy landscape that represents the 
crystallographic states of the 
defect-free (e.g., parent austenite)
and defect (e.g., twinned martensite) lattice,
identifiable from an order parameter; refer \fig{fig2}.
Third, 
a framework for capturing the crystallographic transformation 
in terms of a probabilistic evolution equation of the defect configuration 
on the energy landscape, perturbed by a known time-dependent stimulus, e.g., strain or stress.

We show that the mapping of phase-transformation dynamics as a 
\emph{random-escape} process over the free-energy barrier,  
prescribed within a framework of statistical mechanics, such as the Smoluchowski's equation
can consistently explain the statistical distribution of critical force observed under different strain-rates 
\cite{risken,Hanggi1990,Freund2009,Langer1968}.
Solving this framework is particularly useful as it provides expressions of 
perturbation dependent Kramers \cite{Kramers1940}  escape rates of the parent state over the energy barrier, yielding the rate of 
phase transition as a function of the biasing force.
In addition, expressions of probability fluxes and extant probability 
can be derived, enabling quantitative analyses of the 
dynamic force response patterns. 
In the following sections, the general approach of representing a critical force distribution and a specific model derived using it is described. The utility of the model is demonstrated, and key implications are discussed 
towards gaining a microscopic perspective of the martensitic structural phase transformation from an analysis of 
mechanically induced force responses. 

\section*{Results and Discussions}
\section*{Statistical nature of the mechanically induced structural transition}

Transformation of austenite to martensite, in response to a steadily increasing uniaxial tensile strain, is analysed in a single crystal of titanium nickel alloy from \emph{all-atom} non-equilibrium molecular dynamics simulations \cite{dfrenkel96}. Strain is applied \emph{quasistatically} and homogeneously to a model specimen in the austenite phase, which has a B2 structure \cite{Chowdhury2017}, along the [100] direction according to the protocol: $\Delta l_x(t) = \dot{l_x} t$, where $\Delta l_x(t) = l_x(t) - l_x(0)$ denotes an instantaneous expansion of the box length $l_x$ in the $x$-dimension, and $\dot{l_x}$ is a constant rate of expansion or displacement applied; refer to \fig{fig1}. Maximum strain, $[\Delta l_x(t)/l_x(t=0)]$, applied is restricted to 2.5\%. This regime is characterised by twin-mediated pseudoelastic behaviour. Note that dislocation-slip based plasticity does not occur in the regime simulated.
With progressively increasing strain, a net resistive force $f(t)$ counteracting the deformation, determined from the element, $\sigma_{xx}$, of the internal stress tensor is found to 
develop in the B2 structure (see details under {\bf Methods}). 
The mean constitutive response $\langle f(t) \rangle$ vs.~$\langle \Delta l_x(t) \rangle$, representative 
of a \emph{macroscopic} behaviour is shown, where
$\langle \cdots \rangle$ refers to an ensemble-averaged value, computed as a mean at a given $t$ 
over all the traces and under identical conditions. The decline of average force beyond the linear elastic regime 
of austenite 
correlates with the progress of martensitic transformation \cite{Ye2010,Shaw1997}. 
In what follows, we focus on the force--time traces 
to unravel the characteristics of martensitic transformation. 

\begin{figure}
\centerline{\includegraphics[width=\linewidth,trim= 0mm 0mm 0mm 0mm,clip]{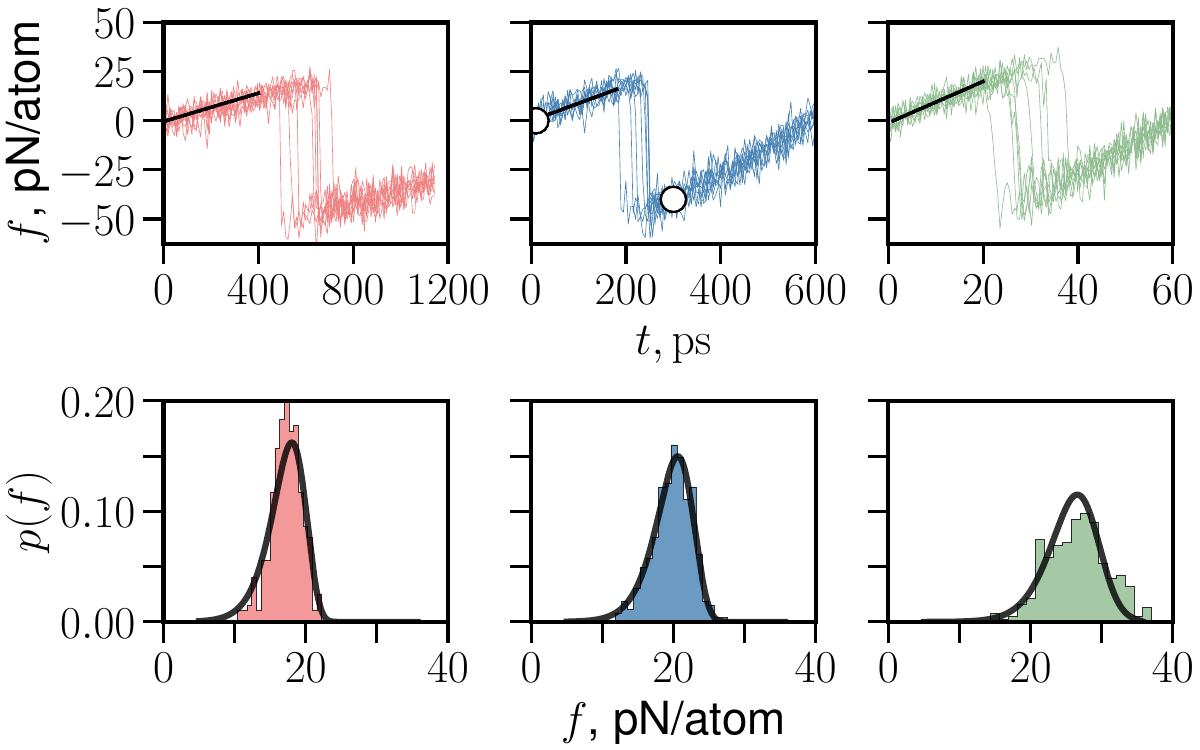}}
\caption{{\bf Stochastic Force--Time Responses. \label{fig3} } \textbf{Top}: Force, $f$, vs.~time sample traces from MD simulations for extension rates of 7.83 $\times$ 10$^{-4}$ (left), 2.35 $\times$ 10$^{-3}$ (middle) and 2.35 $\times$ 10$^{-2}$ \AA/ps (right). Black solid lines are $\langle f(t) \rangle$, and slopes $\partial_t \langle f(t) \rangle $ are the respective force rates, $\dot{f}$ = 3.097 $\times$ 10$^{-2}$, 0.452 $\times$ 10$^{-2}$, and 0.9242 pN/ps per atom for the three cases. White circles mark the times of the snapshots in \fig{fig1}. \textbf{Bottom}: (Color bars) Probability density distribution of transition force, $p(f \, | \, \dot{f})$, corresponding to the peak forces seen in the top-row. Black solid lines are the model predictions using \eqn{fdistr}.}
\end{figure}

\fig{fig3} (Top-Row) depicts representative $f$--$t$ traces acquired from independent and identical MD simulations in response to three  different uniaxial displacement rates (left to right panels): $\dot{l_x}$ = 7.83 $\times$ 10$^{-4}$ \AA/ps, 2.35 $\times$ 10$^{-3}$ \AA /ps, and 2.35 $\times$ 10$^{-2}$ \AA /ps.
%
An individual $f$--$t$ trace increases linearly before
reaching a certain level, $f^*$, which is referred here as \emph{transition} or \emph{critical} force, 
just before falling sharply.
Snapshots of local atomic configurations reveal that $f^*$ corresponds to the onset of martensitic transformation \cite{Shaw1997}.
The peak force has relaxed after stable martensite (product) has formed, 
relieving the stress in the deformed austenite lattice. 

Force--time traces can be treated as signatures reflecting the physical evolution of crystal lattice under mechanical deformation. 
\fig{fig1} (Bottom-Row) depicts the structural differences between initial unstressed austenite 
and post-transformed lattice after the transition is complete. Red and grey dots, which reference the titanium and nickel atoms respectively,  are joined by a bond (line) if a pair of atoms has an interatomic distance less than or equal to 3 \AA. 
Closely packed local directions clarify the visualisation of the lattice restructuring process; a representative movie can be viewed in the SI.

The instantaneous average force, $\langle f(t) \rangle$, grows linearly in time, $t$, associated with the regime of linear elasticity, prior to the occurrence of phase transition; see black lines, \fig{fig3} (Top-Row). 
So, a relation holds:
\begin{equation}
\langle f(t) \rangle = \dot{f} t \equiv f(t) \label{f-t}
\end{equation}
where $\dot{f} = \partial_t \langle f(t) \rangle$ translates into a constant rate of externally applied force prior to the transformation. We will use $\dot{f}$ in the place of $\dot{l_x}$ 
to denote an independent variable \cite{Note3}. 

The $f$--$t$ traces in \fig{fig3} (Top-Row) show that the critical force and the corresponding
onset-time of transformation are statistically distributed.
While a transition force observed in an individual trace is a random variable and cannot be predicted, a histogram of the set $\{ {f_1}^*, {f_2}^*, \cdots, {f_S}^* | \dot{f}\}$ extracted from a large number ($S = 300$) of MD simulations at a given force-rate $\dot{f}$ is well defined; refer to \fig{fig3} (Bottom-Row). 
Further, the mean and standard deviation of the histograms (normalised) are found to trend positively with $\dot{f}$. 
It implies that as the force-rate increases, 
a progressively higher force is needed to reshape 
austenite lattice
because of prior lattice distortion.
The implication is that the observed distribution, 
$p(f^*)$, which embodies microscopic fluctuations during phase transition,  
is an outcome of the kinetic variability of the transformation mechanism, which we model in the next section. 

\section*{Statistical model of a structural transition}

A microscopic process such as the activation of a crystallographic twin involving a cluster of atoms
is intrinsically probabilistic and can be modelled as a random walk.
A description of  a transition process as a random walk is provided by the 
Smoluschowski equation \cite{Garg1995,Evans1999a,Dudko2003,Dudko2006,risken,Hanggi1990}, which captures the evolution
of a probability density function, $P(\xi, t \, | \, \xi=0, t=0)$, for observing the system 
state $\xi$ at time $t$ on an energy landscape $U(\xi, t)$; see \fig{fig2}. 
Here, $\xi$ is a variable denoting an order parameter of the system, 
and we consider $\xi \equiv \Delta l_x$ in the present case. 
When a quasistatic tensile force ramp is applied to the material, the equilibrium energy landscape is progressively deformed
according to
$U(\xi, t) = U_0(\xi) - f(t) \, \xi$, where $U_0(\xi)$ is the equilibrium (no force) free energy profile \cite{Note3}. 
$U_0(\xi)$ is assumed to consist of an attractor \emph{well} domain with its lowest energy state at $\xi = \xi_0$ = 0, and an energy \emph{barrier}, $ \Delta U_\ddag(f=0) = U_0(\xi_\ddag) - U_0(\xi_0)  =  U_\ddag$ located at a \emph{transition} state $\xi = \xi_{\ddag}$. The well is mapped to the austenite (parent) phase, which is entrapped on one side of the barrier. 
To complete the transformation, the entrapped states are required to cross over the 
energy barrier to the other side---martensite phase.
In general, landscape deformation is modelled in terms of the force-dependent landscape features: $\xi_0(f)$, $\xi_\ddag (f)$, and $\Delta U_\ddag(f) = U(\xi_\ddag(f), f) - U(\xi_0(f), f)$, which accelerate the transformation dynamics in the presence of a tensile force.

The equation of motion of $P(\xi, t)$, captured in the Smoluschowski equation, is given as: 
$\partial_t P(\xi, t)  = -\partial_\xi J(\xi, t) $,
where 
$J(\xi, t) = - (k_B T/ \eta) \, \partial_\xi P(\xi, t) + (-\partial_\xi U(\xi, t) / \eta ) \, P(\xi, t)$ 
is a probability flux of the escape of austenitic states by barrier crossing,
and $\eta$ is lattice friction constant of the restructuring of austenite in units of inverse time.
A brief outline of the analytical solutions
\cite{Garg1995,Evans1999a,Dudko2003,Dudko2006,risken,Hanggi1990,Freund2009,Friddle2008a,Maitra2010} is provided below.
Under steady-state flux and an absorbing boundary located at the transition state, the differential equation can be expressed in terms of the survival probability of austenite phase, $\Psi(t) = \int_{-\infty}^{\xi_\ddag} P(\xi, t) d\xi$, and further replacing the variable $t$ by $f$ using \eqn{f-t} gives:
\begin{fleqn}
\begin{align}
&  p(f) = -\frac{d\Psi(f)}{df} = \frac{\Gamma(f) \Psi(f)}{\dot{f} } \label{diffeqn} \\
& \Gamma(f) = \frac{k_BT}{\eta} \left [  \int\limits_{-\infty}^{\xi_\ddag}  d\xi \left\{ e^{-u(\xi, f) } \medint\int_{\xi}^{\xi^+} d\xi_1  e^{u(\xi_1, f)}  \right\} \right ]^{-1} \label{MFPT0}
\end{align}
\end{fleqn} 
Here, $p(f)$ is the transition force distribution, $\dot{f}=\partial_t \langle f(t) \rangle$ is the force rate, 
$u(\xi, f)$ is the free-energy in units of $k_B T$, i.e., $ u(\xi, f) \equiv U(\xi, f)/(k_B T) $, 
and 
$\Gamma(f)$ is a reciprocal of mean passage time to escape the well and provides the force-dependent Kramers escape rate defining the austenite to martensite transition on a time-dependent energy landscape.
The expressions of $p(f)$ and $\Gamma(f)$ facilitate the derivation of parameterized closed-form models
for further analyses of force responses.

Parameterized expressions of $p(f)$ and $\Gamma(f)$ can be derived using an analytical 
free-energy function:
$U_0(\xi) = U_\ddag/2 + (3U_\ddag/2\xi_\ddag) ( \xi -  \xi_\ddag/2) - (2 U_\ddag/(\xi_\ddag)^3) (\xi - \xi_\ddag/2)^3$, which has a form displayed in \fig{fig2} \cite{Note5, Garg1995}.
\eqn{MFPT0} can be simplified 
if $(U_\ddag/k_BT) \gg 1$, and a condition of quasistatic rate of change of the energy landscape, i.e., deformation
applied on a timescale  much longer in comparison to the timescale of phase transition, is assumed. 
These conditions permit the double integral to be expressed as a product of the inner and outer integrals, each evaluated in the subdomains of the well and barrier respectively. 
Substitution of the energy function in \eqn{MFPT0} yields \cite{Garg1995}:
\begin{equation}
\Gamma(f)  \approx \Gamma_0 \left \{ 1 - (f/f_c) \right \}^{1/2} e^{( U_\ddag / k_B T) \left [ 1 - \left \{ 1 - (f/f_c)  \right \}^{3/2} \right ]  }. \label{gammaf}
\end{equation}
where $\Gamma_0 = \{ 1/(2 \pi \eta ) \} \cdot (6U_\ddag/{\xi_\ddag}^2) \;  e^{-U_\ddag/kT}$ is the rate constant at $f=0$ and $f_c \equiv (3U_\ddag/2\xi_\ddag )$ is the maximal force to create martensite.
\eqn{gammaf} shows that the rate of martensitic transformation can be increased exponentially by an applied force $f$. Even small values of force, $f \ll f_c$,  can strongly accelerate the rate of crystallographic twinning 
according to 
$\Gamma(f) \propto \exp (f \xi_\ddag/k_BT)$.

To derive an expression of the probability density distribution of transition 
force, $p(f)$, first, an expression of the survival probability function, $\Psi(f)$, is obtained 
by integrating \eqn{diffeqn}: 
$\int_1^\Psi (\partial \Psi/\Psi) = - [ \int_0^f \Gamma(f) \, \partial f ]/\dot{f} $,
after the substitution of $\Gamma(f)$ from \eqn{gammaf}.
Second, employing the solution of $\Psi(f)$ in \eqn{diffeqn} (first equality) yields \cite{Dudko2006}: 
\begin{equation}
p(f \, | \, \dt{f}) = \frac{\Gamma(f) e^{\mu_0}}{ \dt{f}} \exp { \left \{ - \mu(f)  \left ( 1 - \frac{f}{f_c} \right )^{-1/2} \right \}}, \label{fdistr}
\end{equation}
where $\mu(f) \equiv (\Gamma(f) \, k_B T) / ( \dot{f} \, \xi_\ddag )  $ and $\mu_0 \equiv \mu(f=0)$.
The expression, $p(f \, | \, \dot{f}) \, df$, provides the conditional probability of the austenite to twinned martensite transition at an applied force $f$ and loading rate $\dot{f}$.
\eqn{gammaf} and \eqref{fdistr} are the expressions 
that can be used to retrieve the intrinsic rate of transition ($\Gamma_0$), activation free energy (viz., $U_\ddag$) 
and the interaction range ($\xi_\ddag$).

\begin{figure}
\centerline{\includegraphics[width=\linewidth,trim= 0mm 0mm 0mm 0mm,clip]{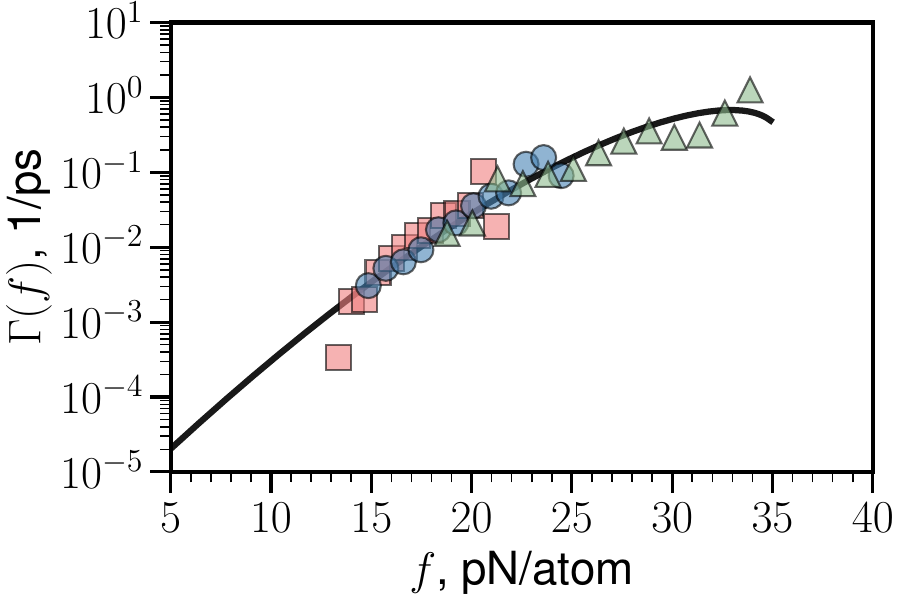}}
\caption{ { {\bf Force Dependent Rate of Martensitic Transition---Force Spectrum. } \label{fig4} }  Symbols are the values of $ \Gamma(f; \dot{f}) \equiv p(f \, | \, \dot{f})/\int_f^\infty df p(f \, | \, \dot{f})$ obtained using the simulation data in \fig{fig3} (Bottom). $\square$, $\iscircle$, $\triangle$ correspond, respectively, to transition rates $\Gamma(f; \dot{f})$ where force rates are $\dot{f}$ = 3.097 $\times$ 10$^{-2}$, 9.452 $\times$ 10$^{-2}$, and 0.9242 pN/ps per atom, as in \fig{fig3}. A solid line is a fit of \eqn{gammaf}.}
\end{figure}

\section*{Energetic and kinetic properties of structural transition}
We evaluate 
the model outlined in the previous section to ascertain its predictive power 
through an analysis of an extensive set of isothermal force responses generated using MD simulations. 
\fig{fig4} shows the force spectra, where the symbols indicate simulation-derived force-dependent rates of phase transition, $\Gamma(f)$, at the three different strain rates. 
These were obtained by converting $p(f \, | \, \dot{f})$, which were extracted from the $f$--$t$ traces and 
shown in \fig{fig3} (Bottom-Row, colour shaded), using 
$\Gamma(f) = \dot{f} p(f \,  | \, \dot{f})/\int_f^{\infty} df p(f \, | \, \dot{f})$ \cite{Dudko2006}, 
a relation derived from \eqn{diffeqn}. 
In addition, a \emph{single} least-squares fit of \eqn{gammaf} was performed on $\Gamma(f)$,
choosing $k_B T = 41.4195$ pN \AA, where $T = 300$ K is the temperature used in the simulations.
The line of best fit is plotted in \fig{fig4}, and 
the best fit parameter-values obtained are $\Gamma_0 = 1.06 \times 10^{-6}$ 1/ps, $U_\ddag = 620$ pN \AA $\approx$ 0.39 eV $\approx$ 15 $k_B T$, and $\xi_\ddag = 26$ \AA. 
The rate constant and activation free energy obtained is found close to the estimates 
reported in Ref.~\cite{Niitsu2020}, however, in our case, the underlying rate process is due to the formation of 
pseudoelastic crystallographic twins, and unrelated to a dislocation-slip based dynamics 
implicated in plastic deformation. 
The predictions, $p(f \, | \, \dot{f})$, after substitution of the extracted parameters in \eqn{fdistr}, are shown as solid lines in \fig{fig3} (Bottom-Row). 

The agreement between models and simulation-derived observables over a large variation of  applied strain rates
shows that the microscopic description based on a non-equilibrium Kramers-Smoluschowski framework used 
for modelling the transition dynamics 
provides a consistent interpretation of the stochasticity observed in the 
force response patterns emerging from the process of martensitic phase transformation. 
While we rely on simulations to generate the mechanical responses, we anticipate realization of probability distributions discussed in this communication is feasible using nanomechanical techniques.
Experimentally obtained distributions can be subsequently interpreted using the theoretical approach for the recovery of mechanistic properties that define the phase transformation.

\section*{Conclusions}
Structural phase transformation in single crystalline metallic systems, triggered by controlled time-dependent 
deformation, reflects in the form of stochastic stress response patterns at the nanoscale.
Such responses can be expressed in terms of statistical distributions. 
It is shown that a distribution function is interpretable
and carries mechanistic information of the phase transition process.
Mapping the phase transformation dynamics to a 
random probabilistic evolution of states over a time-dependent 
free energy barrier is found to reliably elucidate the 
distribution of critical phase-transition force and force-dependent rate 
of transformation, providing an alternative route for 
accessing the otherwise hidden and innate mechanistic properties of a solid-to-solid transformation. 

\section*{Methods}

\paragraph*{\textbf{Molecular Dynamics Simulation:}} 
Nitinol, an alloy of titanium and nickel in equiatomic proportion, was simulated using classical molecular dynamics \cite{dfrenkel96}. The simulation box dimensions used were $l_x$ = 60 \AA,  $l_y$ = 30 \AA \space and $l_z$ = 30 \AA \space along the $x$, $y$ and $z$ axes, aligned respectively to the [100], [010] and [001] crystallographic directions. Periodic boundary conditions were applied on every axis. 
The initial atomic configuration was created using \emph{Atomsk} \cite{HIREL2015212} 
by positioning 2000 atoms, each of titanium and nickel, on the lattice sites of a B2 supercell. 
The B2 unit cell structure (in essence, a bcc lattice) had a lattice parameter of 3 \AA~before equilibration and the basis atoms were placed at (0, 0, 0) and (1/2, 1/2, 1/2) representing nickel and titanium, respectively, as shown in \fig{fig1} (Bottom-Left).

Equilibrium and non-equilibrium molecular dynamics (NEMD) simulations were performed using LAMMPS \cite{Plimpton1995,Guo2017}. 
The interatomic potential employed is the second nearest-neighbour modified embedded-atom method by Ko et al \cite{Ko2015,ctcms}, 
The positions and velocities of the atoms were evolved using 
a timestep of 1 fs. The initial configuration was equilibrated for 1 ns under isothermal and isobaric conditions. Temperature and pressure were constrained using the N\'{o}se-Hoover scheme at \mbox{$T=300$ K} and \mbox{$P = 1.013$ bar}. The damping parameters used for the thermostat and barostat were 0.7 ps and 1 ps, respectively.  

In the  NEMD simulations, the box was deformed at a fixed tensile strain rate along the $x$-direction. 
To ensure that the initial configuration---positions and velocities are distinct and random, an equilibration run of a duration of 1/2 ns preceded every NEMD simulation.
The barostat was turned on only along $y$ and $z$ directions, while the thermostat was active along all axes. 
The simulation box length, $l_x$, was ramped linearly in time according to
$l_x(t) = \dot{l_x} \,  t$.
under the imposed rate of tensile displacement $\dot{l_x}$, which was kept constant. For statistical analyses, $S=300$ simulations were performed for a given displacement rate.  
In all our simulations, a single twinned-sublattice formed. 

The instantaneous resistive force generated per atom in the model system was computed as
$f = \sigma_{xx} \cdot \left( A_{zy} / n \right)$,
where $\sigma_{xx}$ is a normal stress component of the 
internal stress tensor, $(n/A_{zy})$ is the number density of atoms in the $yz$ plane of the simulation box with $n=100$, and $A_{zy} = l_z l_y$ is the cross-sectional area of the $yz$ plane of the simulation box. The ensemble averages of the other elements of the stress tensor were 
approximately zero, showed no evidence of association with the phase transition signatures 
and hence those elements were not considered in the analyses.

\paragraph*{\textbf{Data Analyses:}} 

The raw data comprised of force-rate ($\dot{f}$) specific $f$--$t$ traces, see \fig{fig3} (Top-Row).
The time of occurrence of the peak (or transition) force, just prior to the sharp drop in force level,
was extracted from every trace and enumerated for a given $\dot{f}$ as
$\{\tau_1^*, \tau_2^*, \cdots, \tau_i^*, \cdots \tau_S^* \, | \, \dot{f} \}$,
where $i$ is an index of a simulation trace and $S$ = 300 is the number of MD simulations performed per $\dot{f}$.
The list of times is converted to
$\{f_1^*, f_2^*, \cdots, f_i^*, \cdots f_S^* \, | \, \dot{f} \}$
 via $f_i^* = \dot{f} \tau_i^*$ using \eqn{f-t}, and
transformed further into a \emph{normalized} histogram of phase-transition forces, $p(f \, | \, \dot{f})$; see \fig{fig3} (Bottom-Row).

The force-dependent rate of martensitic transformation, shown as coloured symbols in \fig{fig4}, for a given $\dot{f}$ is computed from the normalised histograms using 
$\Gamma_j(f_j \, | \, \dot{f}) = \dot{f} p(f_j \, | \, \dot{f}) / \sum_j^{n_b} f_j p(f_j \, | \, \dot{f})$, where $\Gamma_j$ is the value of the transition rate at a force $f_j$ corresponding to the $j^{th}$ bin of the histogram, and $n_b = 18$ is the number of bins in the histogram. 

To recover the parameters $U_\ddag$, $\xi_\ddag$ and $\Gamma_0$ of martensitic transformation,
\eqn{gammaf} is fit to the datapoints $\{ \cdots, (f_j, \Gamma_j), \cdots \}$ encompassing all the three force-rates $\dot{f}$ used in this work.
In the fitting procedure, a loss function $L(U_\ddag, \Gamma_0, \xi_\ddag)$, which is a  sum of squared residuals 
\begin{equation}
L(U_\ddag, \Gamma_0, \xi_\ddag) = \sum\limits_{\{\dot{f} \}} \sum\limits_j^{n_b} \left [ \ln  \hat{\Gamma}_j (f_j \, | \, \dot{f})   - \ln(\Gamma_j ) \right ]^2
\end{equation} 
is minimized with respect to the variations of $U_\ddag$, $\Gamma_0$, and $\xi_\ddag$ using \emph{conjugate gradient} algorithm giving a reduced $\chi^2$ value of 0.23 for the best-fit parameters ($\hat{\Gamma}_j$ is the predicted value of transition rate given by \eqn{gammaf} at a force $f_j$).
Note, $\Gamma_j$ corresponding to the tail-regions ($| f - \mu_f | > 2 \sigma_f $) of the normalised histograms $p(f)$, which have only a few samples, were excluded in the fitting process ($\mu_f$ and  $\sigma_f$ denote mean and standard deviation of a normalised histogram).  
Open-source python libraries \emph{pandas}, \emph{matplotlib} and \emph{lmfit} were used for data analysis, charts, and non-linear curve fitting.

\bibliography{main.bbl}

\end{document}